\documentclass[fleqn,a4paper]{vch-book}
\usepackage{amsmath}
\usepackage{amssymb}
\usepackage{makeidx}
 \usepackage{times}
\usepackage{tabularx}
\usepackage{graphicx}

\usepackage{graphics}
\usepackage{booktabs}
\usepackage[dvips]{epsfig}
 \DeclareGraphicsExtensions{.eps,.ps}

 \chardef\bslash=`\\ 
 
 \newcommand{\el}{{\rm el}}
 \newcommand{\dis}{{\rm dis}}

\begin{document}
\setcounter{chapter}{7}
\setcounter{page}{189}
\chapter{Collision of Adhesive Viscoelastic Particles
\for{toc}{\newline}\tocauthor{N. V. Brilliantov and T. P\"oschel}}
\label{chap:BrilliantovPoeschel}
\label{Chapter_Poeschel}
\authorafterheading{Nikolai V. Brilliantov and Thorsten P\"oschel}
\vspace*{-5.5cm} In: H. Hinrichsen and D. Wolf (eds.), {\em The Physics of Granular Media}, Wiley-VCH (Berlin, 2004), p. 189-209\vspace*{3.9cm}


\index{adhesive particles|(}
\index{contact problem}

\subsection*{Abstract}  
The collision of convex bodies is considered for small impact velocity, when plastic deformation and fragmentation may be disregarded. In this regime the contact is governed by forces according to viscoelastic deformation and by adhesion. The viscoelastic interaction is described by a modified Hertz law, while for the adhesive interactions, the model by Johnson, Kendall and Roberts (JKR) is adopted. We solve the general contact problem of convex viscoelastic bodies in quasi-static approximation, which implies that the impact velocity is much smaller than the speed of sound in the material and that the viscosity relaxation time is much smaller than the duration of a collision. We estimate the threshold impact velocity which discriminates restitutive and sticking collisions. If the impact velocity is not large as compared with the threshold velocity, adhesive interaction becomes important, thus limiting the validity of the pure viscoelastic collision model.

\section{Introduction}

The large set of phenomena observed in granular systems, ranging from sand and
powders on Earth to granular gases in planetary rings and protoplanetary
discs, is caused by the specific particle interaction. Besides elastic forces,
common for molecular or atomic materials (solids, liquids and gases),
colliding granular particles also exert dissipative forces. These forces
correspond to the dissipation of mechanical energy in the bulk of the grain
material as well as on their surfaces. The dissipated energy transforms into
energy of the internal degrees of freedom of the grains, that is, the
particles are heated. In many applications, however, the increase in
temperature of the particle material may be neglected (see, e.g.
\cite{BrilliantovPoeschelOUP}).

The dynamical properties of granular materials depend sensitively on the
details of the dissipative forces acting between contacting grains. Therefore,
choosing the appropriate model of the dissipative interaction is crucial for
an adequate description of these systems. In real granular systems the
particles may have a complicated non-spherical shape, they may be non-uniform
and even composed of smaller grains, kept together by adhesion. The particles
may differ in size, mass and in their material properties. In what follows we
consider the contact of granular particles under simplifying conditions. We
assume that the particles are smooth, convex and of uniform material. The
latter assumption allows us to describe the particle deformation by continuum mechanics, disregarding their molecular structure.

It is assumed that particles exert forces on each other exclusively via pairwise mechanical contact, i.e., electromagnetic interaction and gravitational attraction are not considered.

\section{Forces Between Granular Particles} 

\subsection{Elastic Forces}
When particles deform each other due to a static (or quasi-static) contact they experience an elastic interaction force. Elastic deformation implies that, after separation of the contacting particles, they recover their initial shape, i.e., there is no plastic deformation. The stress tensor 
\index{stress tensor}
$\sigma^{ij}_{\rm el} (\vec{r}\,)$ describes the $i$-component of the force, acting on a unit surface which is normal to the direction $j$ ($i,j =\{x,y,z\}$). In the elastic regime the stress 
\index{stress tensor!elastic component} 
is related to the material deformation 
\begin{equation}
\label{u_def}
u_{ij}(\vec{r}\,)=\frac 12 \left( \frac{\partial u_i}{ \partial x_j}  + \frac{\partial u_j}{ \partial x_i} \right) \, ,
\end{equation}
where $\vec{u}(\vec{r}\,)$ is the displacement field at the point $\vec{r}$ in the deformed body, via the linear relation
\begin{equation}
\label{sigma_def_el}
\sigma^{ij}_{\rm el} (\vec{r}\,)=
E_1 \left(u_{ij}(\vec{r}\,) -\frac13 \delta_{ij} u_{ll}(\vec{r}\,) \right) +E_2 \delta_{ij} u_{ll}(\vec{r}\,)\,.
\end{equation}
Repeated indices are implicitly summed over (Einstein convention). The coefficients $E_{1}$ and $E_2$ read
\begin{equation}
\label{E1E2}
E_1=\frac{Y}{(1+\nu)}, \quad \qquad E_2= \frac{Y}{3(1-2\nu)} \,,
\end{equation}
where $Y$ is the Young modulus 
\index{Young module}
and $\nu$ is the Poisson ratio.
\index{Poisson ratio}
Let the pressure $\vec{P}(x,y)$ act on the surface of an elastic semispace, $z>0$, leading to a displacement field in the bulk of the semispace \cite{Landau:1965}:
\begin{equation}
\label{ui_Pk}
u_i= \iint G_{ik}\left( x-x^{\prime}, y-y^{\prime},z  \right) P_k \left(x^{\prime}, y^{\prime} \right) dx^{\prime} d y^{\prime} \,,
\end{equation}
where $G_{ik}\left( x, y, z \right)$ is the corresponding Green function. For the contact problem addressed here we need only the $z$-component of the displacement on the surface $z=0$, that is, we need only the component
\begin{equation}
\label{Green_zz}
G_{zz}(x,y,z=0)=\frac{(1-\nu^2)}{\pi  Y}\frac{1}{\sqrt{x^2+y^2}} = \frac{(1-\nu^2)}{\pi  Y}\frac{1}{r}
\end{equation}
of the Green function \cite{Landau:1965}.

Consider a contact of two convex smooth bodies labeled as 1 and 2. We assume that only normal forces, with respect to the contact area, act between the particles. In the contact region their surfaces are flat. For the coordinate system centered in the middle of the contact region, where $x=y=z=0$, the following relation holds true:
\begin{equation}
\label{x2y2uz1uz2}
B_1x^2 +B_2y^2 +u_{z1}(x,y)+u_{z2}(x,y)=\xi \,,
\end{equation}
where $u_{z1}$ and $u_{z2}$ are respectively the $z$-components of the displacement in the material of the first and of the second bodies  on the plane $z=0$. The sum of the compressions of both bodies in the center of the contact area 
\index{contact area}
defines $\xi$. The constants $B_1$ and $B_2$ are related to the radii of curvature of the surfaces in contact \cite{Landau:1965}:
\begin{equation}
  \label{B1B2_R1R2}
  \begin{split}
    2\left( B_1+B_2\right) = & \frac{1}{R_1}+\frac{1}{R_2}+\frac{1}{R_1^{\prime}} + \frac{1}{R_2^{\prime}} \\ 
\!\!\!\!\!\!\!\!\!\!    4\left( B_1\!-\!B_2\right)^2 = &\!  \left(\! \frac{1}{R_1}\!-\!\frac{1}{R_2}\!\right)^2 \!\!+ \!\left(\!\frac{1}{R_1^{\prime}} \!-\!\frac{1}{R_2^{\prime}}\right)^2 \!\!+ 2 \cos 2 \varphi\!  \left(\! \frac{1}{R_1}\!-\!\frac{1}{R_2}\right)\left(\!\frac{1}{R_1^{\prime}}\! -\!\frac{1}{R_2^{\prime}} \right) . 
\end{split}
\end{equation}
Here $R_1$, $R_2$ and $R_1^{\prime}$, $R_2^{\prime}$ are respectively the
principal radii of curvature of the first and the second body at the point of
contact and  $\varphi$ is the angle between the planes corresponding to the
curvature radii $R_1$ and  $R_1^{\prime}$. Equations~(\ref{x2y2uz1uz2}), (\ref{B1B2_R1R2}) describe the general case of the contact between two smooth bodies (see \cite{Landau:1965} for details). The physical meaning of \eqref{x2y2uz1uz2} is easy to see for the case of a contact of a soft sphere of a radius $R$ ($R_1=R_2=R$) with a hard, undeformed  plane ($R_1^{\prime}=R_2^{\prime}=\infty$). In this case $B_1=B_2=1/R$, the compressions of the sphere and of the plane are respectively $u_{z1}(0,0)=\xi$ and $u_{z2}=0$, and the surface of the sphere before the deformation is given by $z(x,y)=(x^2+y^2)/R$. Then \eqref{x2y2uz1uz2} reads in the flattened area $u_{z1}(x,y)= \xi - z(x,y)$, that is, it gives  the condition for a point $z(x,y)$ on the body's surface to touch the plane $z=0$. 

The displacements  $u_{z1}$ and $u_{z2}$ may be expressed in terms of the normal pressure $P_z(x,y)$ which acts between the compressed bodies in the plane $z=0$. Using \eqref{ui_Pk} and \eqref{Green_zz} we rewrite \eqref{x2y2uz1uz2}  as
\begin{equation}
\label{Eq_for_uz12}
\frac{1}{\pi} \left( \frac{1-\nu_1^2}{Y_1} +  \frac{1-\nu_2^2}{Y_2} \right) \int\!\!\! \int \frac{P_z(x^{\prime},y^{\prime})}{r}dx^{\prime}d y^{\prime} = \xi -B_1x^2-B_2y^2 \,,
\end{equation}
where $r=\sqrt{(x- x^{\prime})^2+ (y- y^{\prime})^2}$ and integration is performed over the contact area. 
Equation \eqref{Eq_for_uz12} is an integral equation for the unknown function $P_z(x,y)$. We compare this equation with the mathematical identity \cite{Landau:1965}
\begin{equation}
\label{identity}
\!\!\!\int\!\!\! \int \!\!\frac{dx^{\prime}d y^{\prime}}{r} \sqrt{1\!-\!\frac{x^{\prime\, 2}}{a^2} \!-\!\frac{y^{\prime \, 2}}{b^2}}= 
\frac{\pi ab}{2} \!\int\limits_0^{\infty} \left[\!  1\!-\!\frac{x^2}{a^2+t}\!-\!\frac{y^2}{b^2+t} \right] \,,
\!\frac{dt}{\sqrt{(a^2+t)(b^2+t)t}} 
\end{equation}
where integration is performed over the elliptical area $x^{\prime\, 2}/a^2+y^{\prime\, 2}/b^2=1$. The left-hand sides of both equations contain integrals of the same type, while the right-hand sides contain quadratic forms of the same type. Therefore, the contact area is an ellipse with the semi-axes $a$ and $b$ and the pressure is of the form $P_z(x,y) ={\rm const} \sqrt{1-x^2/a^2-y^2/b^2}$. The constant may be found from the total elastic force $F_\el$ acting between the bodies. Integrating $P_z(x,y)$ over the contact area we obtain
\begin{equation}
\label{Pressure}
P_z(x,y) =\frac{3F_\el}{2\pi ab} \sqrt{1-\frac{x^2}{a^2} -\frac{y^2}{b^2}} \,.
\end{equation}  
We substitute \eqref{Pressure} into \eqref{Eq_for_uz12} and replace the double integration over the contact area by integration over the variable $t$, according to the identity \eqref{identity}. Thus, we obtain an equation containing terms proportional to $x^2$, $y^2$ and a constant.
Equating the corresponding coefficients we obtain
\begin{eqnarray}
\label{xi_Fel}
&&\xi=\frac{F_\el D}{\pi} \int_0^{\infty} \frac{dt}{\sqrt{(a^2+t)(b^2+t)t}} =\frac{F_\el D}{\pi} \frac{N(x)}{b}\\
\label{B1_Fel}
&&B_1=\frac{F_\el D}{\pi} \int_0^{\infty} \frac{dt}{(a^2+t) \sqrt{(a^2+t)(b^2+t)t}} 
=\frac{F_\el D}{\pi} \frac{M(x)}{a^2b}\\
\label{B2_Fel}
&&B_2=\frac{F_\el D}{\pi} \int_0^{\infty} \frac{dt}{(b^2+t) \sqrt{(a^2+t)(b^2+t)t}} \, 
=\frac{F_\el D}{\pi} \frac{M(1/x)}{ab^2}\,,
\label{B3x}
\end{eqnarray}
where  
\begin{equation}
\label{def_D}
D \equiv \frac34 \left( \frac{1-\nu_1^2}{Y_1} +  \frac{1-\nu_2^2}{Y_2} \right)
\end{equation}
and $x \equiv  a^2/b^2$ is the ratio of the contact ellipse semi-axes. In \eqref{xi_Fel}-\eqref{B3x} we introduce the short-hand notations\footnote{The function $N(x)$ and $M(x)$ may be expressed as a combination of the Jacobian elliptic functions $E(x)$ and $K(x)$ \cite{AbramowitzStegun:1965}.}
\begin{eqnarray}
\label{eq:def_N}
&&N(x)=\int_0^{\infty} \frac{dt}{\sqrt{(1+xt)(1+t)t}} \\
\label{eq:def_M}
&&M(x)= \int_0^{\infty} \frac{dt}{(1+t) \sqrt{(1+t)(1+xt)t}} \, .
\end{eqnarray}
From these relations will follow the size of the contact area, $a$, $b$, and the compression $\xi$ as functions of the elastic force $F_\el$ and the geometrical coefficients $B_1$ and $B_2$. 

The dependence of the force $F_\el$ on the compression $\xi$ may be obtained from scaling arguments. If we rescale $a^2 \to \alpha a^2$, $b^2 \to \alpha b^2$, $\xi  \to \alpha \xi$ and $F_\el \to \alpha^{3/2} F_\el$, with  $\alpha$ constant, Eqs.~(\ref{xi_Fel})--(\ref{B2_Fel}) remain unchanged. That is, when $\xi$ changes by the factor $\alpha$, the semi-axis $a$ and $b$ change by the  factor $\alpha^{1/2}$ and the force by the  factor  $\alpha^{3/2}$, i.e., $a \sim \xi^{1/2}$, $b \sim \xi^{1/2}$ and 
\begin{equation}
\label{Fel_xi_gen}
F_\el = {\rm const } \, \xi^{3/2} \, .
\end{equation}
The  dependence \eqref{Fel_xi_gen} holds true for all smooth convex bodies 
\index{convex bodies}
in contact. To find the constant in \eqref{Fel_xi_gen} we divide \eqref{B2_Fel} by \eqref{B1_Fel} and obtain the transcendental equation 
\begin{equation}
\label{eq:k}
\frac{B_2}{B_1}= \frac{\sqrt{x} M \left(1/x \right)}{M(x)}
\end{equation}
for the ratio of semi-axes $x$. Let $x_0$ be the root of Eq. \eqref{eq:k}, then $a^2=x_0b^2$ and we obtain 
\begin{eqnarray}
\label{eq:xi_N0}
&&\xi=\frac{F_\el D}{\pi}\frac{N(x_0)}{b} \\
\label{eq:B1_M0}
&&B_1=\frac{F_\el D}{\pi}\frac{M(x_0)}{x_0b^3} \, , 
\end{eqnarray}
where $N(x_0)$ and $M(x_0)$ are pure numbers. Equations~(\ref{eq:xi_N0}),
(\ref{eq:B1_M0}) allow us to find the semi-axes $b$ and the elastic force $F_\el$ as functions of the compression $\xi$. Hence we obtain the force, i.e., we get  the  constant in \eqref{Fel_xi_gen}:
\index{elastic interaction force}
\begin{equation}
\label{Fel_xi_gen1}
F_\el = \frac{\pi}{D} \left( \frac{M(x_0)}{B_1 x_0 N(x_0) } \right)^{1/2} \, \xi^{3/2}
= C_0 \xi^{3/2}\,.
\end{equation}
For the special case of contacting spheres ($a=b$), the constants $B_1$ and $B_2$ read 
\begin{equation}
\label{Spheres}
B_1=B_2=\frac12 \left( \frac{1}{R_1}+\frac{1}{R_2} \right)=\frac12 \frac{1}{R^{\rm eff}} \, .
\end{equation}
In this case $x_0=1$, $N(1)=\pi$, and $M(1)=\pi/2$, leading to the solution of
(\ref{eq:xi_N0}), (\ref{eq:B1_M0}): 
\index{Hertz contact law}
\begin{eqnarray}
\label{eq:a2_Hertz}
&&a^2=R^{\rm eff} \, \xi \\
\label{Fel_xi_sph}
&&F_\el = \rho \xi^{3/2} \, ;  \qquad \rho \equiv \frac{2Y}{3(1-\nu^2)} \sqrt{R^{\rm eff}} \, , 
\end{eqnarray}
where we use the definition \eqref{def_D} of the constant $D$. 
This contact problem was solved by Heinrich Hertz in 1882 \cite{Hertz:1882}. It describes the force between {\em elastic} particles. For inelastically deforming particles it describes the repulsive force in the static case. 

\subsection{Viscous  Forces}
\index{viscous interaction force}

When the contacting particles move with respect to each other, i.e., the
deformation changes with time, an additional dissipative force arises, which
acts in the opposite direction to the relative particle motion. The dissipative processes occurring in the bulk of the body cause a viscous contribution to the stress tensor. For small deformation the respective component of the stress tensor is proportional to the deformation rate $\dot{u}_{ij}(\vec{r}\,)$, according to the general relation \cite{BrilliantovSpahnHertzschPoeschel:1994}:
\begin{equation}
\label{Vis_str_gen}
\!\!\!\!\! \sigma^{ij}_\dis (\vec{r},t)=
E_1\!\! \int\limits_0^t\!\! d\tau \psi_1(t-\tau)\! \left[\dot{u}_{ij}(\vec{r},\tau) -\frac13 \delta_{ij} \dot{u}_{ll}(\vec{r},\tau) \right]
\!+E_2\!\! \int\limits_0^t\!\! d\tau \psi_2(t-\tau) \delta_{ij} \dot{u}_{ll}(\vec{r},\tau) \, , 
\end{equation}
where the (dimensionless) functions $\psi_1(t)$ and $\psi_2(t)$ are the relaxation functions for the distortion deformation and 
$\psi_2(t)$ for the dilatation deformation. 

In many important applications the viscous stress tensor 
may be simplified significantly. If the relative velocity of the colliding bodies is much smaller than the speed of sound in the particle material and if the characteristic relaxation times of the dissipative processes $\tau_{\rm vis, \, 1/2}$ are much smaller than the duration of the collision $t_c$,
\begin{equation}
\label{tau1tau2}
\tau_{\rm vis, \, 1/2} \equiv \int_0^{\infty}  \psi_{1/2}(\tau) d\tau \ll t_c \, , 
\end{equation}
the viscous constants $\eta_1$ and $\eta_2$ may be used instead of the functions $\psi_1(t)$ and $\psi_2(t)$. Thus
\begin{equation}
\label{eta1eta2}
\eta_{1/2}=E_{1/2} \tau_{\rm vis, \, 1/2} = E_{1/2} \int_0^{\infty}  \psi_{1/2}(\tau) d\tau \, 
\end{equation}
and the dissipative stress tensor reads (see \cite{BrilliantovSpahnHertzschPoeschel:1994} for details)
\begin{equation}
\label{Vis_Str}
\sigma^{ij}_\dis (\vec{r},t)=\eta_1 \left[\dot{u}_{ij}(\vec{r},\tau) -\frac13 \delta_{ij} \dot{u}_{ll}(\vec{r},\tau) \right] + \eta_2 \delta_{ij} \dot{u}_{ll}(\vec{r},\tau) \, .
\end{equation}
It may be also shown that the above conditions are equivalent to the assumption of 
\index{quasi-static deformation}
quasi-static deformation \cite{BrilliantovSpahnHertzschPoeschel:1994,BrillSpaHerPoesPhysA:1996}. When the material is deformed quasi-statically, the displacement field $\vec{u}(\vec{r}\,)$ in the particles coincides with that for the static case $\vec{u}_\el(\vec{r}\,)$, which is the solution of the elastic contact problem. The field $\vec{u}_\el(\vec{r}\,)$, in its turn, is completely determined by the compression $\xi$, which varies with time during the collision, i.e., $\vec{u}_\el=\vec{u}_\el(\vec{r}, \xi)$. Therefore, the corresponding displacement rate may be approximated as
\begin{equation}
\label{udot_xidot}
\dot{\vec{u}}(\vec{r},t) \simeq \dot{\xi} \frac{\partial }{\partial \xi} \vec{u}_\el(\vec{r}, \xi) \, 
\end{equation}
and  the dissipative stress tensor reads, respectively
\begin{equation}
\label{Vis_Str_dotxi}
\sigma^{ij}_\dis = \dot{\xi} \frac{\partial }{\partial \xi}  \left[ 
\eta_1 \left(u_{ij}^\el -\frac13 \delta_{ij}u_{ll}^\el\right) +
\eta_2 \delta_{ij} u_{ll}^\el \right] \, .
\end{equation}
From \eqref{Vis_Str_dotxi} and \eqref{sigma_def_el} follows the relation between the elastic and dissipative stress tensors within the quasi-static approximation, 
\index{stress tensor!dissipative component}
\begin{equation}
\label{ST_elas_dis}
\sigma^{ij}_\dis = \dot{\xi} \frac{\partial }{\partial \xi} \sigma^{ij}_\el
\left(E_1 \leftrightarrow \eta_1, E_2 \leftrightarrow \eta_2 \right)  \, ,
\end{equation}
where we emphasize that the expression for the dissipative tensor may be obtained from the corresponding expression for the elastic tensor after substituting the elastic constants by the relative viscous constants, and application of the operator $\dot{\xi} \partial / \partial \xi$.

The component $ \sigma^{zz}_\el$ of the elastic stress is equal to the normal pressure $P_z$ at the plane $z=0$ of the elastic problem, Eq. \eqref{Pressure}
\begin{equation}
\label{sigma_zz_Pz}
\begin{split}
\sigma^{zz}_\el(x,y,0)=& E_1 \frac{\partial u_z}{\partial z} + \left(E_2-\frac{E_1}{3} \right)
\left(\frac{\partial u_x}{\partial x} +\frac{\partial u_y}{\partial y}+\frac{\partial u_z}{\partial z} \right) \\
=&\frac{3F_\el}{2 \, \pi ab} \sqrt{1-\frac{x^2}{a^2} -\frac{y^2}{b^2}} \,.
\end{split}
\end{equation}
Now we compute the total dissipative force acting between the bodies. Instead of a direct computation of the dissipative stress tensor, we employ the method  proposed in \cite{BrilliantovSpahnHertzschPoeschel:1994,BrillSpaHerPoesPhysA:1996}: We transform the coordinate axes as
\begin{equation}
\label{xyzx1y1z1}
x= \alpha x^{\prime}\,,\hfill 
y= \alpha y^{\prime}\,,\hfill 
z= z^{\prime}\hfill
\end{equation}
with 
\begin{align}
\label{alpbetab}
\alpha&=\left(\frac{\eta_2-\frac13 \eta_1}{\eta_2+\frac23 \eta_1} \right) 
\left(\frac{E_2+\frac23 E_1}{E_2-\frac13 E_1} \right) &
\beta&=\frac{(\eta_2-\frac13 \eta_1)}{\alpha( E_2-\frac13 E_1)} \\
a&=\alpha a^{\prime} &
b&=\alpha b^{\prime} \, .
\end{align}
and perform the transformations
\begin{equation}
\label{eta12_to_E12}
\begin{split}
\eta_1 &\frac{\partial u_z}{\partial z} + \left(\eta_2-\frac{\eta_1}{3} \right) 
\left(\frac{\partial u_x}{\partial x} +\frac{\partial u_y}{\partial y}+\frac{\partial u_z}{\partial z} \right)
 \\
&=\beta \left[
E_1 \frac{\partial u_z}{\partial z^{\prime}} + \left(E_2-\frac{E_1}{3} \right) 
\left(\frac{\partial u_x}{\partial x^{\prime}} +\frac{\partial u_y}{\partial y^{\prime}}+\frac{\partial u_z}{\partial z^{\prime}} \right)\right]
\\
&=\beta \frac{3F_\el}{2 \, \pi a^{\prime}b^{\prime}} \sqrt{1-\frac{x^{\prime \, 2}}{a^{\prime \, 2}} -\frac{y^{\prime \, 2}}{b^{\prime \, 2}}}
=\beta \alpha^2 \frac{3F_\el}{2 \, \pi ab} \sqrt{1-\frac{x^2}{a^2} -\frac{y^2}{b^2}} \, .
\end{split}
\end{equation}
Applying the operator $\dot{\xi} \partial / \partial \xi$ to the last expression on the right-hand side we obtain the dissipative stress tensor. Subsequent integration over the contact area yields, finally, the total dissipative force acting between the bodies:
\begin{equation}
\label{Fdis_Fel1}
F_\dis=A \dot{\xi} \frac{\partial} {\partial \xi} F_\el (\xi) \, , 
\end{equation}
where 
\begin{equation}
  \label{eq:Adef}
  A\equiv \alpha^2 \beta =\frac13\frac{\left(3\eta_2-\eta_1\right)^2}
{\left(3\eta_2+2\eta_1 \right)}
\left[\frac{\left(1-\nu^2\right)(1-2\nu)}{Y\nu^2}\right] \, .
\end{equation}
 Using the scaling relations for the elastic force, Eq. \eqref{Fel_xi_gen},  and for the semi-axes of the 
\index{contact ellipse}
contact ellipse, we obtain
\begin{equation}
\label{eq:dF_da_db_dxi}
\frac{\partial F_\el}{\partial \xi}=\frac{3}{2}\frac{F_\el}{\xi} \, , \qquad
\frac{\partial a }{\partial \xi}=\frac{1}{2}\frac{a}{\xi} \, , \qquad
\frac{\partial b }{\partial \xi}=\frac{1}{2}\frac{b}{\xi} \, . 
\end{equation}
Then from \eqref{eta12_to_E12} and \eqref{Fel_xi_gen1},  the distribution of the dissipative pressure in the contact area 
\index{contact area}
may be found:
\begin{equation}
\label{Diss_Press}
P_{z}^\dis (x,y) = \frac{3 A}{4 \pi} \, \frac{AC_0}{ab} \dot{\xi} \sqrt{\xi}
\, \left(1-\frac{x^2}{a^2} -\frac{y^2}{b^2}\right)^{-1/2} \, ,
\end{equation}
where the constant $C_0$ is defined in \eqref{Fel_xi_gen1}. 

We wish to stress that, to derive  the above expressions, we  assumed only
that  the surfaces of the two bodies in the vicinity of the contact point
before the deformation, are described by  the quadratic forms
$z_1=\kappa_{ij}^{(1)}x_ix_j$ and $z_2=\kappa^{(2)}x_ix_j$ ($i,j=x,y,z$),
where  $ \kappa^{(1/2)}_{ij}$ are symmetric tensors
\cite{Landau:1965}. Therefore, the  relations obtained are  valid for a contact of arbitrarily shaped convex bodies.
For spherical particles of identical material, \eqref{Fdis_Fel1} and \eqref{Fel_xi_sph} yield \cite{BrilliantovSpahnHertzschPoeschel:1994,BrillSpaHerPoesPhysA:1996}
\begin{equation}
\label{eq:Fdis_Sph}
F_\dis= \frac32 A \rho \dot{\xi} \sqrt{\xi}  \, ,  
\end{equation}  
with $\rho$ as defined in \eqref{Fel_xi_sph}. Hence, the total force acting between viscoelastic spheres takes the simple form \cite{BrilliantovSpahnHertzschPoeschel:1994,BrillSpaHerPoesPhysA:1996}
\begin{equation}
\label{eq:Fdis_SphA}
F=\rho \left( \xi^{3/2} +\frac32 A \sqrt{\xi} \dot{\xi} \right)  \, . 
\end{equation}  
The range of validity of \eqref{eq:Fdis_SphA} for the viscoelastic force is determined by the quasi-static approximation. The impact velocity must be significantly smaller than the speed of sound. On the other hand, the impact velocity must not be too small in order to neglect adhesion. We also neglect plastic deformation in the material.

\subsection{Adhesion of Contacting Particles}
\index{adhesion}

\subsubsection{Models of Adhesive Interaction}

The Hertz theory has been derived for the contact of non-adhesive particles. Adhesion becomes important when the distance of the particle surfaces approaches the range of molecular forces. Johnson, Kendall and Roberts (JKR) \cite{JKR:1971} 
\index{JKR theory}
extended the Hertz theory by taking into account adhesion in the flat contact region. They show that the contact area is enlarged by the action of the adhesive force. Therefore, they introduced an apparent Hertz load $F_H$ which would cause this enlarged contact area. To simplify the notation, we consider the contact of identical spheres. The contact area is then a circle of radius $a$, which corresponds to the compression $\xi_H$ for the Hertz load $F_H$. In reality, however, this contact radius occurs at the compression $\xi$ which is smaller than $\xi_H$. In the JKR theory it is assumed that the difference between the Hertz compression $\xi_H$ and the actual one, $\xi$, may be attributed to the additional stress
\begin{equation}
\label{eq:P_B}
P_B(x,y)=\frac{F_B}{2 \pi a^2} \, \left( 1-\frac{r^2}{a^2}\right)^{-1/2}  \, ,
\end{equation}
which is the solution of the classical Boussinesq problem \cite{Timoshenko:1970}. This distribution of the normal surface traction gives rise to  a constant displacement over a circular region of an elastic body. The displacement $\xi_B$ corresponding to the contact radius $a$ and the total load $F_B$ are related~by
\begin{equation}
\label{eq:xi_B}
\xi_B= \frac23 D \frac{F_B}{a} \, ,
\end{equation}
where the constant $D$ is defined in \eqref{def_D}.

The value of $F_B<0$ mimics  the additional surface forces, such that the pressure is positive (compressive) in the center of the contact area, while it is negative (tensile) near the boundary \cite{JKR:1971}. Hence, the shape of the body is determined by the action of two effective forces $F_H$ and $F_B$. The total force between the particles is their difference, $F=F_H-F_B$. Johnson \textit{et al.} assumed that the elastic energy stored in the deformed spheres may be found as a difference of the elastic energy corresponding to the Hertz force $F_H$ and that due to the force $F_B$ \cite{JKR:1971}.  Using
\begin{equation}
\label{eq:contEner}
U_s=-\pi \gamma a^2     
\end{equation}
for the surface energy, where  $\gamma >0$ is twice the surface free energy
per unit area of the solid in vacuum or gas, and minimizing the total energy,
we find  \cite{JKR:1971}
\begin{equation}
\label{eq:F_B}
F_B=-2\pi a^2 \sqrt{\frac{3 \gamma}{2 \pi D a}} \, ,
\end{equation}
and, thus, the contact radius corresponding to the total force $F$:
\begin{equation}
\label{eq:ContRad}
a^3=\frac12 D \, R \left( F+ \frac32 \pi \gamma R +\sqrt{3 \pi \gamma R F + \left(\frac32 \pi \gamma R \right)^2 } \right)
\end{equation}  
and also the compression
\begin{equation}
\label{eq:compres_ad}
\xi=\frac{2a^2}{R} - \sqrt{\frac{8 \pi \gamma D a}{3}}\, .
\end{equation}
The first term in \eqref{eq:compres_ad} is the Hertz compression $\xi_H$, which coincides with \eqref{eq:a2_Hertz} for $R^{\rm eff} \to R/2$. Equation \eqref{eq:ContRad} may be solved to express  the total force as a function of the contact radius: 
\begin{equation}
\label{eq:Ftot_via_a}
F(a)=\frac{2a^3}{DR}-\sqrt{\frac{6\pi \gamma}{D}} a^{3/2}\, .
\end{equation}
For vanishing applied load the contact radius $a_0$ is finite:
\begin{equation}
\label{eq:a_zero_load}
a_0^3= \frac32 D \pi \gamma R^2 \, .
\end{equation} 
For negative applied load the contact radius decreases and the condition for a real solution of \eqref{eq:ContRad} yields the maximal negative force which the adhesion forces can resist,
\begin{equation}
\label{eq:F_max}
F_{\rm sep} = - \frac34  \pi \gamma R \,,
\end{equation}
corresponding to the contact radius
\begin{equation}
\label{eq:a_sep}
a_{\rm sep}^3 = \frac38 D \pi \gamma R^2 = \frac14 a_0^3 \, .
\end{equation}
For a larger (in the absolute value) negative force, the spheres separate. For spheres of dissimilar radii, in (\ref{eq:ContRad})--(\ref{eq:a_sep}) $R$ should be substituted by $2 R^{\rm eff}$.


Another approach to the problem of the adhesive contact was developed by
Derjaguin, Muller and Toporov (DMT). They  assumed that the Hertz profile of
the pressure  distribution on the surface stays unaffected by adhesion and
obtained the pull-off force $F_{\rm sep} = - 2 \pi \gamma R^{\rm eff}$
\cite{DerjaguinMullerToporov:1975}. The  assumption of the Hertz profile
allows one to avoid the singularities of the pressure distribution \eqref{eq:P_B} on the boundary of the contact zone. Since the experimental measurement of $\gamma$ is problematic, it is not possible to check the validity of the JKR and DMT theories, i.e., to resolve their disagreement. 

In later studies \cite{MullerYuschenkoDerjaguin:1980,MullerYuschenkoDerjaguin:1983} a more accurate theoretical analysis has been performed. The elastic equations have been solved numerically for a simplified microscopic model of adhesive surfaces with Lennard--Jones interaction. 
\index{Lennard--Jones interaction}
Within this microscopic approach, the relative accuracy of different theories has been estimated for a wide range of model parameters. It was found that the DMT theory is valid for small adhesion and for small, hard particles. JKR theory is more reliable for large, soft particles with large adhesion forces, which, however, should be short-ranged.

In \cite{AttardParker:1992} the Lennard--Jones continuum model of solids was
studied. The adhesive forces between the surfaces  then read
\begin{equation}
\label{eq:LJ_surf}
P_s(h)=\frac{H}{6 \pi h^3}\left[ \frac{z_0^6}{h^6}-1 \right] \,.
\end{equation}
Here $P_s(h)$ describes the forces acting per unit area between the surfaces, $h=h(r)$ is the actual microscopic distance between them. $H$ is the Hamaker constant, characterizing the van der Waals attraction of the particles in a gas or vacuum and $z_0$ is the equilibrium separations of the surfaces. The surface energy in this model is defined by
\begin{equation}
\label{eq:gamma_Hammak}
\gamma=\frac{H}{16 \pi z_0^2} \, . 
\end{equation} 
It was observed in \cite{AttardParker:1992} that the accuracies of different theories vary depending of the value of the Tabor parameter $\mu$, 
\index{Tabor parameter}
\cite{Tabor:1997}
\begin{equation}
\label{eq:Tabor_par}
\mu^{3/2} \equiv \frac23 \gamma D \sqrt{R^{\rm eff}/z_0^3} \, .
\end{equation} 
In agreement with
\cite{MullerYuschenkoDerjaguin:1980,MullerYuschenkoDerjaguin:1983} it has been
shown \cite{AttardParker:1992} that small values of $\mu$ (small hard
particles with low surface energies) favor the DMT theory ($\mu < 10^{-2}$)
while for $\mu \sim 1$--$10$ the JKR theory proves to be rather more accurate. Both JKR and DMT fail for large $\mu$  when the strong  adhesion is combined with the soft material of the contacting bodies. In this limit, the surfaces jump into contact, which corresponds to a spontaneous non-equilibrium transition (see e.g. \cite{SmithBozzoloetal:1989}). Similar analysis has been performed later \cite{Greenwood:1977}, where the author concluded that the DTM theory generally fails, both in original and corrected forms. One of the main conclusions of \cite{AttardParker:1992,Greenwood:1977} is that the JKR theory, albeit simple,  gives relatively accurate predictions for basic quantities  in the range of its validity ($\mu \sim 1$--$10$).

Among the theories developed to cover the DMT--JKR transition regimes \cite{MullerYuschenkoDerjaguin:1980,MullerYuschenkoDerjaguin:1983,Greenwood:1977,Tabor:1997,HughesWhite:1980,Maugis:1992} the theory by Maugis \cite{Maugis:1992} is the most frequently used. It is based on  a simplified model of adhesive forces. The adhesive force of constant intensity $P_D$ is extended over a fixed distance $h_D$ above the surface, yielding the surface tension $\gamma=P_D h_D$. The description of a contact in this model is based on two coupled analytical equations which are to be solved numerically. The recently developed double-Hertz model \cite{GreenwoodJohnson:1998,Haiatetal:2003} constructs the solution for the adhesive contact as a sum of two Hertzian solutions, which make the theory analytically more tractable than the Maugis model. Combining, in the adopted manner, the successful assumptions of the JKR and the modified DMT model, a generalized analytical theory for the adhesive contact has been proposed \cite{Schwarz:2003}.

In what follows we assume that the parameters of our system belong to the range of validity of the JKR model, $\mu \sim 1 - 10$,  
and will use this simple analytical theory to describe the adhesive contacts between spheres. Moreover, we assume that the adhesive force is small. To estimate the influence of the adhesive force, we approximate $\xi \approx 2a^2/R$ in  \eqref{eq:compres_ad}  and substitute it into \eqref{eq:Ftot_via_a} to obtain (see also \cite{SpahnAlbersetal:2004}), 
\begin{equation}
\label{eq:FtotJKR}
F \approx \rho \xi^{3/2} - \sqrt{6 \pi \gamma / D} \left(  R^{\rm eff} \right)^{3/4} \xi^{3/4} \, .
\end{equation}

\subsubsection{Viscoelasticity in Adhesive Interactions}

The adhesive forces between particles cause the additional deformation in the contacting bodies as compared to a pure Hertzian deformation, hence in the corresponding dynamical problem an additional deformation rate arises. Therefore, the dissipative forces must have an additional component attributed to the adhesive interactions. The adhesive contact of viscoelastic spheres has been studied numerically in \cite{SpahnAlbersetal:2004,Haiatetal:2003}. In \cite{Brilliantovetal:2004} the quasi-static condition for the colliding  viscoelastic adhesive spheres was used and an analytical expression for the interaction force has been derived for the JKR model. Similar to the case for non-adhesive particles, it was assumed that in the quasi-static approximation,  the deformation field may be parameterized by the value of the compression $\xi$. (Note that this assumption neglects the possible hysteresis which can happen for the negative total force \cite{AttardParker:1992}). Performing the same transformation which lead to the expression \eqref{Fdis_Fel1} for the case of non-adhesive contact,  and using the approximation \eqref{eq:FtotJKR} we obtain the estimate for the dissipative forces \cite{Brilliantovetal:2004}
\index{dissipative force}
\begin{eqnarray}
\label{eq:Fdis_adh}
&&F_\dis= \frac32 A \rho \dot{\xi} \sqrt{\xi}  + \frac34 B \sqrt{6 \pi \gamma / D} \left(  R^{\rm eff} \right)^{3/4}  \dot{\xi} \xi^{-1/4} \\
\label{eq:B_adh}
&&B  \equiv \alpha \beta = \frac{(3 \eta_2 -\eta_1) Y \nu}{3(1+\nu)(1-2 \nu)} \,.
\end{eqnarray}
Note the singularity in the second term of \eqref{eq:Fdis_adh} at $\xi=0$ \footnote{This is a weak or integrable singularity, that is $\int_0^{\epsilon} \xi^{-1/4} d\xi \sim \epsilon^{3/4} \to 0$ for $\epsilon\to 0 $. Hence for practical application of \eqref{eq:Fdis_adh} one can use $\xi > \epsilon$, where $\epsilon$ may be very small but a finite number.}.  It is attributed  to the quasi-static approximation for JKR theory and physically reflects the fact that the adhesive particles can jump  into contact \cite{SmithBozzoloetal:1989} with the discontinuous change of the compression $\xi$.
Consider now how the above forces determine the particle dynamics.

\section{Collision of Granular Particles}
\index{collision of particles}

\subsection{Coefficient of Restitution}
\index{coefficient of restitution}
\index{restitution coefficient|\see coefficient of restitution}

\begin{figure}[htbp]
  \centerline{\includegraphics[width=4cm,clip]{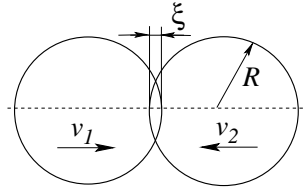}}
  \caption{Head-on collision of identical spheres. The time-dependent state is characterized by the compression $\xi(t) \equiv 2R-|\vec{r}_1(t)-\vec{r}_2(t)|$ and the compression rate  $\dot{\xi}(t)=v_1(t)-v_2(t)$.}
  \label{fig:psiatcollis}
\end{figure}

Based on the particle interaction forces discussed so far, we turn now to the
description of the particle collisions. It is assumed that the colliding
particles do not exchange tangential forces\footnote{See
  \cite{BrilliantovPoeschelOUP} for a discussion of the consistency of this
  assumption.}, hence, only normal motion is considered. Let the particles be
spheres of the same material, which start to collide at time $t=0$ at relative
normal velocity $g$ (impact rate). The time-dependent compression  then reads
\begin{equation}
\label{eq:xi_R1R2r1r2}
\xi(t)= R_i+R_j-\left|\vec{r}_i(t) -\vec{r}_j(t)\right| \, ,
\end{equation} 
where $\vec{r}_i(t)$ and $\vec{r}_j(t)$ are positions of the particle centers at time $t$ (see Figure~\ref{fig:psiatcollis}).
The relative normal motion of particles at a collision is equivalent to the motion of a point particle with the effective mass
\begin{equation}
\label{eq:eff_mass}
m^{\rm eff}=\frac{m_im_j}{m_i+m_j} \, .
\end{equation} 
For the moment let us neglect adhesion and consider the collision of viscoelastic particles interacting via the force \eqref{eq:Fdis_SphA}. The equation of motion and the initial conditions read
\begin{equation}
\label{eq:eom}
\ddot{\xi} + \frac{\rho}{m^{\rm eff}} \left( \xi^{3/2} +\frac32 A \sqrt{\xi} \dot{\xi} \right) =0\,,\hfill
\dot{\xi}(0)=g\,,\hfill
\xi(0)=0 \,.\hfill
\end{equation}
When granular particles collide, part of the energy of the relative motion is dissipated. The coefficient of (normal) restitution quantifies this phenomenon:
\begin{equation}
  \label{eq:edefratio}
  \varepsilon=-\dot{\xi}\left(t_c \right) / \dot{\xi}(0)
=-\dot{\xi}(t_c)/g\,,
\end{equation}
where $\dot{\xi}(0)=g$ is the pre-collision relative velocity and $t_c$ is the duration of the collision. In general, $\varepsilon$ is a function of the impact velocity. It can be obtained by integrating \eqref{eq:eom} numerically \cite{HertzschSpahnBrilliantov:1995,BrilliantovSpahnHertzschPoeschel:1994,BrillSpaHerPoesPhysA:1996} or analytically \cite{SchwagerPoeschel:1998}.

\subsection{Dimensional Analysis}

The analytical solution \cite{SchwagerPoeschel:1998} requires considerable efforts, here we give a simplified derivation which is based on a dimensional analysis of the equation of motion \eqref{eq:eom} \cite{Ramirez:1999}. This method was employed before \cite{Tanaka} to prove that the frequent assumption $\varepsilon=$const. is inconsistent with mechanics of materials.
For the dimensional analysis, the elastic and dissipative forces are represented in the more general form
\begin{equation}
  \label{eq:ForcesGenForm}
  F_{\rm el} = m^{\rm eff} D_1 \,\xi^{\alpha}\,,\hfill F_{\rm diss} = m^{\rm eff} D_2\, \xi^{\gamma}\dot\xi^{\beta}\,, \hfill
\end{equation}
with $D_{1/2}$ being material parameters. With these notations the
equation of motion for colliding particles reads
\begin{equation}
    \ddot \xi +D_1\, \xi^{\alpha} + D_2\, \xi^{\gamma}\,\dot {\xi}^{\beta} =0\,,\hfill
    \dot{\xi}(0)= g\,,\hfill
    \xi(0) = 0\,.\hfill
\end{equation}
For the case of pure elastic deformation ($D_2=0$) the maximal compression $\xi_0$ is obtained by equating the initial kinetic energy, $m^{\rm eff} g^2/2$ and the elastic energy $m^{\rm eff} D_1 \xi_0^{\alpha+1}/(\alpha+1)$:
\begin{equation}
\xi_0 \equiv \left
(\frac{\alpha+1}{2\, D_1} \right )^{1/(1+\alpha)}
{g}^{2/(1+\alpha)}\,.
\end{equation}
We chose $\xi_0$ as the characteristic length of the problem. The time needed
to cover the distance $\xi_0$ when traveling at velocity $g$ defines the characteristic time: $\tau_0\equiv \xi_0/g$. Thus, the dimensionless variables read
\begin{equation}
  \label{eq:dimlessvars}
  \hat{\xi} \equiv \xi/\xi_0 \,, \hfill
  \dot{\hat{\xi}} \equiv \dot\xi/g \,, \hfill
  \ddot{\hat{\xi}} = \left(\xi_0 /g^2 \right) \ddot\xi \,.\hfill
\end{equation}
In dimensionless form, \eqref{eq:eom} reads
\begin{equation}
\label{eq:eomdimles}
\ddot{\hat{\xi}} + \varkappa \, \hat{\xi}^{\gamma}\,\dot{\hat{\xi}}{}^{\beta} +\frac{1+\alpha}{2}\,\hat{\xi}^{\alpha} =0\,,\hfill
\dot{\hat{\xi}}(0)=1\,,\hfill
\hat{\xi}(0) = 0\hfill
\end{equation}
with
\begin{equation}
\label{eq:deltadef}
\varkappa= \varkappa(g) = 
D_2\, \left (\frac{1+\alpha}{2\, D_1}\right )^{(1+\gamma)/(1+\alpha)}\,
g^{2 (\gamma -\alpha)/(1+\alpha)+\beta}\,.
\end{equation}
None of the terms in \eqref{eq:eomdimles} depends either on material properties or on impact velocity, except for $\varkappa$. Therefore, if the motion of the particles depends on material properties and on impact velocity, it may depend only via $\varkappa$, i.e., in the combination of the parameters as given by \eqref{eq:deltadef}. Hence, any function of the impact velocity, including the coefficient of restitution must be of the form $\varepsilon(g) = \varepsilon[\varkappa(g)]$. A similar result for $\varepsilon \to 0$, $\beta=1$ and $\alpha=3/2$ has been obtained in \cite{Falconetal:1998}.


Hence, if the coefficient of restitution does not depend on the impact
velocity $g$, it is implied that
\begin{equation}
2(\gamma -\alpha)+\beta\,\left( 1+\alpha\right ) =0\,.
\label{eq:condition}
\end{equation}
For small $\dot{\xi}$ a linear dependence of the dissipative force on the velocity seems to be realistic, i.e., $\beta=1$. Then $\varepsilon=$const. holds true for the following cases: 
\begin{itemize}
\item For the linear elastic force $F_\el \propto \xi$, (i.e. $\alpha=1$) condition \eqref{eq:condition} implies the linear dashpot force  $F_\dis \propto \dot{\xi}$, ($\gamma=0$).

\item For  the Hertz law for 3D-spheres \eqref{Fel_xi_sph}, (i.e. $\alpha=3/2$), condition \eqref{eq:condition} requires $F_\dis \propto \dot\xi\, \xi^{1/4}$, ($\gamma=\frac14$).  As far as we can see there is no physical argument to justify this functional form of the dissipative force.
\end{itemize}
Therefore, we conclude that the assumption $\varepsilon=$const. is in
agreement with mechanics of materials only in the case of (quasi-)one-dimensional systems. For three-dimensional spheres it disagrees with basic mechanical laws. 

For viscoelastic spheres, according to \eqref{eq:Fdis_SphA}, the coefficients
are $\alpha=3/2$, $\beta=1$, and $\gamma=1/2$. From \eqref{eq:deltadef} it
follows that
\begin{equation}
  \label{eq:delta}
    \varkappa=\frac{3}{2}\left(\frac{5}{4}\right)^{3/5} A\left( \frac{\rho} m^{\rm eff}\right)^{2/5}g^{1/5} \, 
\end{equation}
and, therefore, 
\begin{equation}
    \varepsilon =\varepsilon \left[A \left(\frac{\rho} m^{\rm eff} \right)^{2/5} g^{1/5} \right]\,.
\end{equation}
If we assume that the function $\varepsilon(g)$ is sufficiently smooth and can be expanded into a Taylor series, and with $\varepsilon(0)= 1$, for small impact velocity the coefficient of restitution reads
\begin{equation}
  \label{eq:epsguess}
\varepsilon=1-C_1 A \kappa^{2/5} g^{1/5} + C_2 A^2 \kappa^{4/5} g^{2/5} \mp \dots \, 
\end{equation}
where 
\begin{equation}
  \label{eq:defkappa}
\kappa = \left(\frac32 \right)^{5/2} \left(\frac{\rho}{m^{\rm eff}}\right) =
\left(\frac32 \right)^{3/2} \frac{Y\,\sqrt{R^{\rm eff}}}{m^{\rm eff}\left(1-\nu^2\right) }\,.
\end{equation}
The coefficients $C_1$, $C_2, \ldots$ are pure numbers which are given
analytically in \cite{SchwagerPoeschel:1998}.  Here we give a simple derivation of these coefficients (which is correct for $C_1$ and $C_2$ and approximately correct for $C_3$ and $C_4$, using the method proposed in \cite{Ramirez:1999}).

\subsection{Coefficient of Restitution for Spheres}
\subsubsection{Small Inelasticity Expansion}
Using $d/ d\xi = \dot{ \hat{\xi} } d/d \hat{ \xi}$ the equation of motion for a collision adopts the form
\begin{equation}
  \label{eq:Efirst}
\frac{d}{d \hat{\xi}} \left( \frac12 \dot{\hat{\xi}}^2 + \frac12 \hat{\xi}^{5/2} \right) = -\varkappa \dot{\hat{\xi}} 
\sqrt{\hat{\xi}} =\frac{dE(\hat{\xi})}{d\hat{\xi}}\,,\hfill
 \hat{\xi}(0)=0\,,\hfill
\dot{\hat{\xi}} (0) =1\,,\hfill
\end{equation}
where we introduce the mechanical energy
\begin{equation}
E = \frac12 \dot{\hat{\xi}}^2 + \frac12 \hat{\xi}^{5/2}\,.
\end{equation}
The first stage of the collision starts at $\hat{\xi}=0$ and ends in the turning point of maximal compression $\hat{\xi}_0$. During the second stage, the particles return to $\hat{\xi}=0$. The energy dissipation during the first stage is given by
\begin{equation}
\label{eq:Elosdir}
\int_0^{\hat{\xi}_0} \frac{dE}{d\hat{\xi}} d\hat{\xi}= 
-\varkappa \int_0^{\hat{\xi}_0} \dot{\hat{\xi}} \sqrt{\hat{\xi}}
d\hat{\xi} \,.
\end{equation}
For the evaluation of the right-hand side of \eqref{eq:Elosdir}, the
dependence $\dot{\hat{\xi}}=\dot{\hat{\xi}}( \hat{\xi})$ is needed. In the
case of an elastic collision where the maximal compression is $\hat{\xi}_0=1$
(according to the definition of the dimensionless variables) from energy
conservation, it follows that
\begin{equation}
\dot{\hat{\xi}} (\hat{\xi})=\sqrt{1- \hat{\xi}^{5/2}} \, , 
\end{equation}
i.e., $\dot{\hat{\xi}}$ vanishes at the turning point $\hat{\xi}=1$.  For inelastic collisions $\hat{\xi}_0\lesssim 1$, therefore, 
\begin{equation}
\label{eq:xx1}
\dot{\hat{\xi}} (\hat{\xi}) \approx 
\sqrt{1- (\hat{\xi}/\hat{\xi}_0)^{5/2}}\,.
\end{equation} 
Using \eqref{eq:xx1} the integration in \eqref{eq:Elosdir} may be performed yielding
\begin{equation}
\label{eq:direct}
\frac12 \hat{\xi}_0^{\,5/2}-\frac12 = -\varkappa\, b\, \hat{\xi}_0^{\,3/2}
\end{equation} 
where we take into account 
\begin{equation}
E(\hat{\xi}_0)  =\frac12 \hat{\xi}_0^{5/2}\,,\hfill
E(0)=\frac12 \dot{\hat{\xi}}^2(0)=\frac12\hfill
\end{equation}
and introduce the constant
\begin{equation}
  b \equiv \int_0^1 \sqrt{x} \sqrt{1-x^{5/2}}\, dx =
  \frac{ \sqrt{\pi}\, \Gamma \left(3/5\right)}{5 
    \, \Gamma \left(21/10\right)}\,.
  \label{eq:defofd}
\end{equation} 
Let us define  the {\em inverse collision}, the collision that
starts with velocity $\varepsilon \, g$ and ends with velocity $g$. During the
inverse collision, the system gains energy. The maximal compression $\hat{\xi}_0$ is naturally
the same for both collisions, since the inverse collision equals the direct collision, except for the 
fact that time runs in the reverse direction, hence, 
\begin{equation}
\frac{dE(\hat{\xi})}{d\hat{\xi}} =+\varkappa \dot{\hat{\xi}} \sqrt{\hat{\xi}}\,,\hfill
\dot{\hat{\xi}}(0)=\varepsilon \,,\hfill    
 \hat{\xi}(0)=0\,.\hfill
\end{equation}
This suggests an approximative relation for the inverse collision,
\begin{equation}
\dot{\hat{\xi}} (\hat{\xi}) \approx \varepsilon \,
\sqrt{1- (\hat{\xi}/\hat{\xi}_0)^{5/2}}\,,
\end{equation} 
with the additional pre-factor $\varepsilon$, which is the initial velocity for
the inverse collision.

Integration of the energy {\it gain} for the first phase of the inverse collision (which equals, up to its sign, the energy loss in the second phase of the direct collision \cite{SchwagerPoeschel:1998}) may be performed just in the same way as for the direct collision, yielding
\begin{equation}
\label{eq:inverse}
\frac12 \hat{\xi}_0^{5/2}-\frac{\varepsilon^2}{2} = +\varepsilon \, \varkappa\, b\, 
\hat{\xi}_0^{3/2}\,,
\end{equation} 
where again $E(\hat{\xi}_0)=\hat{\xi}_0^{5/2}/2$ and $E(0)=\varepsilon^2/2$ is used.
Multiplying \eqref{eq:direct} by $\varepsilon$ and summing it with \eqref{eq:inverse}, the maximal compression is $ \varepsilon =  \hat{\xi}_0^{\,5/2}$. Substituting this into (\ref{eq:direct}) we arrive at an equation for the coefficient of restitution 
\begin{equation}
\label{eq:eqforeps}
\varepsilon+2\varkappa\, b\, \varepsilon^{3/5}=1\,.
\end{equation} 
The formal solution to this equation may be written as a continuous fraction (which does not diverge in the limit $g\to\infty$):
\begin{equation}
\label{eq:eqforeps1}
\varepsilon^{-1} = 1+ 2\varkappa\, b (1+ 2\varkappa\, b(1+ \cdots)^{2/5} \cdots ) ^{2/5}
\end{equation} 
For practical applications the series expansion of $\varepsilon$ in terms of $\varkappa$ is more appropriate. We return to dimensional units and define the characteristic velocity $g^*$ such that
\begin{equation}
\label{deltaviagstar}
\varkappa \equiv \frac{1}{2b} \left ( \frac{g}{g^*} \right )^{1/5}\,,
\end{equation}
with $b$ being defined in \eqref{eq:defofd}. Using, moreover, the definition \eqref{eq:deltadef} together with \eqref{eq:Fdis_SphA}, which provides the values of $D_1$ and $D_2$, the characteristic velocity reads
\begin{equation}
\label{eq:GDEF}
  \left(g^{*}\right)^{-1/5}=\frac{\sqrt{\pi}}{2^{1/5}5^{2/5}} \frac{\Gamma\left(3/5\right)}{\Gamma\left(21/10\right)} \left(\frac{3}{2}A\right)\left(\frac{\rho}{ m^{\rm eff}}\right)^{2/5}\,.
\end{equation}
With this new notation the coefficient of restitution adopts a simple form:
\begin{equation}
\label{eq:eqforeps2x}
\varepsilon=1-a_1\left(\frac{g}{g^*}\right)^{1/5}\!\!+a_2 
\left(\frac{g}{g^*}\right)^{2/5}\!\! -a_3
\left(\frac{g}{g^*}\right)^{3/5}\!\! + 
a_4 \left(\frac{g}{g^*}\right)^{4/5}\!\! \mp \cdots\,,
\end{equation}
with $a_1=1$, $a_2=3/5$, $a_3= 6/25 =0.24$, $a_4=7/125=0.056$. Rigorous but elaborated calculations \cite{SchwagerPoeschel:1998} show that, while the coefficients $a_1$ and $a_2$ are exact, the correct coefficients $a_3$ and $a_4$ are: $a_3 \approx 0.315$ and $a_4 \approx 0.161$. The coefficients
$C_i$ of the expansion \eqref{eq:epsguess} can be obtained via
\begin{equation}
\label{eq:coeffepsC1}
C_i=a_i C_1^i =a_i \left(g^{*}\right)^{-i/5}\, . 
\end{equation}
In particular, 
\begin{equation}
\label{eq:coeffepsC1C2}
    C_1=\frac{\sqrt{\pi}}{2^{1/5}5^{2/5}} \frac{\Gamma\left(3/5\right)}{\Gamma\left(21/10\right)}\,,\hfill
    C_2=\frac35 C_1^2\,\hfill
\end{equation}
and respectively, $C_3 \approx -0.483$, $C_4 \approx 0.285$. The convergence of the series is rather
slow, and accurate results can be expected only for small enough $g/g^*$.

Let us briefly mention a complication of the 
\index{quasi-static approximation}
quasi-static approximation (QSA). During the expansion phase it may happen that the repulsive force according to \eqref{eq:Fdis_SphA} becomes negative, i.e., seemingly the particles attract each other. For the interaction of non-cohesive particles we had, however, excluded attractive forces. This is an artefact, since in reality the particles lose contact already, before completely recovering their spherical shape, i.e., before $\hat{\xi}=0$ (see \cite{PoeschelSchwagerAlgo} for a detailed explanation of this problem). This effect, however, is not in agreement with the QSA.  Obviously, \eqref{eq:Fdis_SphA} which is a result of the QSA, derived in \cite{BrilliantovSpahnHertzschPoeschel:1994}, is not appropriate to describe the very end of the particle contact. Taking this effect into account we obtain a larger coefficient of restitution as compared with the presented computation \cite{SchwagerPoeschelEpsCorr:2004}. For small dissipation, the correction is rather small. This small correction is neglected here.

\subsubsection{Pad\'e Approximation} 
\index{coefficient of restitution!Pad\'e approximation}

\begin{figure}[tb]
\centerline{\includegraphics[angle=0,width=6cm,clip]{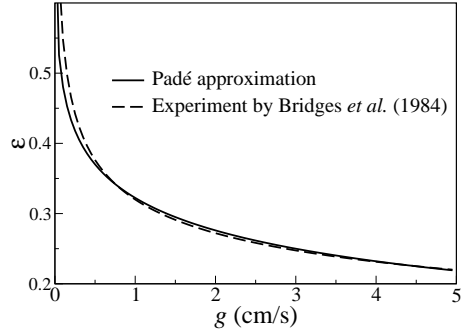}}  
  \caption{Dependence of the coefficient of normal restitution  on the 
    impact velocity for ice particles. The dashed line is experimental \cite{BridgesHatzesLin:1984}, the
    solid line is the
    Pad\'e-approximation \eqref{eq:Padegen} with the constants given
    by \eqref{eq:tabled1d5} and with the characteristic velocity
    for ice $g^*=0.32 \,$~cm\,s$^{-1}$.}
  \label{fig:Bridgescompare}
\end{figure}

For practical applications, such as molecular dynamics simulations, the expansion \eqref{eq:eqforeps2x} is of limited value, since it diverges for large impact velocities, $g \to \infty$. It is possible, however, to construct a Pad\'e approximant for $\varepsilon$, based on the above coefficients, which reveals the correct limits, $\varepsilon(0)=1$ and $\varepsilon (\infty)=0$. The dependence of $\varepsilon(g) $ is expected to be a smooth monotonically decreasing function, which suggests that the order of the numerator must be smaller than the order of the denominator. The 1-4 Pad\'e-approximant 
\begin{equation}
\label{eq:Padegen}
\varepsilon=\frac{1+d_1\left(g/g^*\right)^{1/5}}{1+d_2
\left(g/g^*\right)^{1/5}+d_3\left(g/g^*
\right)^{2/5}+d_4\left(g/g^*\right)^{3/5}+d_5 \left(g/g^*\right)^{4/5}} \,
\end{equation}
satisfies these conditions. Standard analysis (e.g. \cite{Pade}) yields the coefficients $d_k$ in terms of the coefficients $a_k$
\begin{align}
\label{eq:tabled1d5}
&d_0 = a_4-2a_3-a_2^2+3a_2-1  \\
&d_1 =\left[1-a_2+a_3-2a_4+(a_2-1)(3a_2-2a_3)\right]/d_0              &\approx 2.583\nonumber\\
&d_2 =\left[(a_3-a_2)(1-2a_2)-a_4\right]/d_0                          &\approx 3.583\nonumber\\
&d_3 =\left[ a_3+a_2^2(a_2-1)-a_4(a_2+1)\right]/d_0                   &\approx 2.983\nonumber\\
&d_4 =\left[ a_4(a_3-1)+(a_3-a_2)(a_2^2-2a_3)\right]/d_0              &\approx 1.148\nonumber\\
&d_5 = \left[ 2(a_3-a_2)(a_4-a_2a_3)-(a_4-a_2^2)^2-a_3(a_3-a_2^2)\right]/d_0\!\!\!\!\!\!\!\!\!\!\!\!\!\!\!\!\!\!\!\!\!\!\!\!\!\!&\approx 0.326 \nonumber
\end{align}
Using the characteristic velocity $g^*=0.32$~cm\,s$^{-1}$ for ice at very low temperature as a fitting parameter, we compare the theoretical prediction of $\varepsilon (g)$, given by \eqref{eq:Padegen}, with the experimental results \cite{BridgesHatzesLin:1984}, see Figure~\ref{fig:Bridgescompare}.   
The discrepancy with the experimental data at small $g$ follows from the fact
that the extrapolation expression, $\varepsilon =0.32/g^{0.234}$ used by
\cite{BridgesHatzesLin:1984} to fit the experimental data has an unphysical
divergence at $g \to 0$ and does not imply the failure of the theory for this
region. The scattering of the experimental data presented by
\cite{BridgesHatzesLin:1984} is large for small impact velocity, according to
experimental complications, therefore the fit formula of
\cite{BridgesHatzesLin:1984} cannot be expected to be accurate enough for
velocities that are too small. For very high velocities the effects, such as
brittle failure, fracture and others, may contribute to the dissipation, so
that the mechanism of the viscoelastic losses could not be the primary one. In
the region of very small velocity, other interactions than viscoelastic ones, e.g., adhesive interactions, may be important. 

\subsection{Coefficient of Restitution for Adhesive Collisions}
\index{coefficient of restitution!adhesive particles}

For very small velocities, when the kinetic energy of the relative motion of colliding particles is comparable with the surface interaction energy at the contact, the adhesive forces play an important role in collision dynamics -- they may change  the coefficient of restitution qualitatively. Indeed, as described above, adhesive particles in contact are  compressed even for vanishing external load, i.e., a tensile force must be applied to separate the particles. Therefore, at the second stage of the collision, the separating particles must overcome a barrier due to the attractive interaction, which keeps them together. The work against this tensile force reduces the kinetic energy of the relative motion after the collision, that is, it reduces the effective coefficient of restitution. For small impact velocity the kinetic energy of the relative motion may be too small to overcome the attractive barrier, i.e., the particles stick together after the collision, corresponding to $\varepsilon=0$.
\index{sticking collision}

From these arguments it follows that the description of particle collisions by pure viscoelastic interaction has a limited range of validity, not only for large impact rate when plastic deformation becomes important, but also for small impact rata due to adhesion. A simplified analysis of adhesive collisions is presented in \cite{Brilliantovetal:2004} to estimate the influence of adhesive forces on the coefficient of restitution. It allows to estimate the range of validity of the  viscoelastic collision  model. 

We assume that  the JKR theory is adequate for the given system parameters.  We also assume that the adhesion is small and that the adhesive interactions may be neglected when the force between the particles is purely repulsive. Hence, we take into account the influence of adhesive interaction only when the total force is attractive, that is when the force is mainly determined by adhesion. This happens at the very end of the collision. We also neglect the additional dissipative forces, which arise due to the adhesive interaction and assume that all dissipation during the collision may be attributed to the viscoelastic interactions.

At the second stage of a collision, when the particles move away from each other they pass the point where the contact area is $a_0$ and the total force vanishes. As the particles move away further, the force becomes negative, until it reaches, at $a=a_{\rm sep}$, the maximum negative value $F=F_{\rm sep}$, Eq. \eqref{eq:F_max}. At this point the contact of the particles is terminated and they separate. According to our assumption, the work of the tensile force which acts against the particles, separation reads
\begin{equation}
\label{eq:work_tensil}
W_0=\int_{\xi(a_0)}^{\xi(a_{\rm sep})} F(\xi) d\xi = \int_{a_0}^{a_{\rm sep}} F(a) \frac{d\xi}{da} da \, .
\end{equation}
Using \eqref{eq:Ftot_via_a} for the total force $F(a)$, \eqref{eq:compres_ad}
for the compression, which allows one to obtain $d\xi /da$, and
(\ref{eq:a_zero_load}), (\ref{eq:a_sep}) for $a_0$ and $a_{\rm sep}$, we obtain the work of the tensile forces,
\begin{equation}
\label{eq:W_0}
W_0=q_0 \left(  \pi^5 \gamma^5 D^2 R^4\right)^{1/3} \,,
\end{equation}
with the constant
\begin{equation}
\label{eq:def_q1}
q_0=\frac{1}{10} \left( 2^{1/3}\, 3-1 \right)3^{2/3}\,.
\end{equation}
Close to the end of the collision, just before the tensile forces start to act, the relative velocity is $g^{\prime} = \varepsilon g$. The final velocity $g^{\prime \prime}$, when the particles completely separate from each other, may be found from the conservation of energy:
\begin{equation}
\label{eq:gprpr}
\frac12 m^{\rm eff} \left( g^{\prime } \right)^2
-\frac12 m^{\rm eff} \left( g^{\prime \prime} \right)^2  = W_0 \, . 
\end{equation}
From the latter equation we obtain the coefficient of restitution for the adhesive collision, $\varepsilon_{\rm ad}$,
\begin{equation}
\label{eq:eps_adh}
\varepsilon_{\rm ad} (g) = \frac{g^{\prime \prime}}{g} =
\frac{\sqrt{ \varepsilon^2(g)g^2 - 2W_0/m^{\rm eff}}}{g} \, , 
\end{equation}
where $\varepsilon(g)$ is the coefficient of restitution without the adhesive interaction. Hence we obtain the condition for the validity of the viscoelastic collision model,
\begin{equation}
\label{eq:condi_visc}
\varepsilon(g) g  \gg \sqrt{\frac{2 W_0}{m^{\rm eff}}}\, . 
\end{equation}
The threshold impact velocity $g_{\rm st}$  which separates the restitutive ($g>g_{\rm st})$ from the sticking ($g<g_{\rm st}$) collisions, may be obtained from the solution of the equation
\begin{equation}
\label{eq:condi_stick}
 \frac12 m^{\rm eff}\varepsilon^2(g)g^2 = W_0\, .  
\end{equation}
Using \eqref{eq:epsguess} we obtain for viscoelastic spheres, in the leading-order approximation, with respect to the small dissipative parameter $A$:
\begin{equation}
\label{eq:g_stick}
g_{\rm st}= \sqrt{ \frac{2W_0}{m^{\rm eff}} } \left[1+C_1 A \kappa^{2/5} \left( \frac{2W_0}{m^{\rm eff}} \right)^{1/10} \right]\, .
\end{equation}
For head-on collisions (vanishing tangential component of the impact velocity) the colliding particles stick together if $g<g_{\rm st}$. In this case, after the collision, the particles form a joint particle of mass $m_1+m_2$. 

\section{Conclusion}  

We have considered the collision of particles in granular matter with respect
to viscoelastic and adhesive interaction. Thus, the elastic contribution due
to the classical Hertz theory is complemented by the simplest model for
dissipative material deformation, where the viscous stress is linearly related
to the strain rate. Moreover, quasi-static approximation was assumed, i.e.,
the impact velocity is much smaller than the speed of sound in the material
and the viscosity relaxation time is much smaller than the duration of the
collision. Using these approximations, we obtained the general solution for
the contact problem for convex viscoelastic bodies. The validity of this model
is violated for large impact velocity due to plastic deformations and also for
very small impact velocity due to surface forces. We have discussed the
available models of adhesive interaction. For the model by Johnson, Kendall
and Roberts \cite{JKR:1971} which has been shown to be accurate in a range of
parameters of practical interest, the additional dissipation arising due to
adhesive forces have been estimated. From the comparison of the force
contribution due to pure viscoelastic interaction and the contribution due to
adhesion, we have estimated the range of validity of the viscoelastic
model. For head-on collisions we have also estimated the marginal value of the impact velocity, which discriminates restitutive and sticking collisions.

\renewcommand\bibname{References}

\index{adhesive particles|)}

\end{document}